Debjoy Thakur—Dr. Ishapathik Das
Department of Mathematics and Statistics
Indian Institute of Technology Tirupati
Tirupati, Andhra Pradesh, India-517506.
E-mail: debjoythakur95@gmail.com
           ishapathik@iittp.ac.in

# Statistical assessment of spatio-temporal impact of lockdown on air pollution using different modelling approaches in India.

**Abstract**

One of the main contributors to air pollution is particulate matter ($PM_{xy}$), which causes several COVID-19 related diseases such as respiratory problems and cardiovascular disorders. Therefore, the spatial and temporal trend analysis of particulate matter and the mass concentration of all aerosol particles $\leq 2.5\ \mu m$ in diameter ($PM_{2.5}$) has become critical to control the risk factors of co-morbidity of a patient. Lockdown plays a significant role in maintaining COVID-19 cases as well as air pollution, including particulate matter. This study aims to analyse the effect of the lockdown on controlling air pollution in metropolitan cities in India through various statistical modelling approaches. Most research articles in the literature assume a linear relationship between responses and covariates and take independent and identically distributed error terms in the model, which may not be appropriate for analysing such air pollution data. In this study, we performed a pattern analysis of daily $PM_{2.5}$ emissions in various major activity zones during 2019 and 2020. By measuring the lockdown effect, we also considered seasonal influence. We predicted the $PM_{2.5}$ values in the unobserved location using three popular spatial interpolation techniques: (i) inverse distance weight (IDW), (ii) ordinary kriging (OK), and (iii) random forest regression kriging (RFK), and compared their root mean square error (RMSE). Subsequently, we estimated the spatio-temporal intervention of lockdown on air pollution using the difference-in-difference (DID) estimator. In winter, the transport zones, namely Anand Vihar and ITO airport, were the most affected regions. The northwestern part of Delhi is the most sensitive zone in terms of air pollution. Due to the lockdown, the weekly $PM_{2.5}$ emission decreased by 62.15%, the mass concentration of all aerosol particles $\leq 10\ \mu m$ in diameter ($PM_{10}$) decreased by 53.14%, and the air quality index (AQI) decreased by 22.40%. We propose adopting remedies to maintain the air pollution index considering the spatial and temporal variability in the responses.

## 1. Introduction

Air pollution is the biggest threat to public health and is one of the primary causes of respiratory hazards and chronic obstructive pulmonary disease (COPD) (Hernandez 2012). Hence, the risk of a patient being infected by COVID-19 is high. Air pollution has life-threatening consequences on morbidity, which has been discovered by utilising the risk of morbidity due to air pollution (Ri-Map) model (Nagpure 2014). In developing countries, $PM_{10}$ is positively related to urbanisation (Fotourehchi 2016). For example, in Tehran, rapid urban expansion increases nitrogen dioxide (NO2) emissions and other pollutant concentrations in the air, endangering human life (Ghalehteimouri et al. 2021). The Bayesian Model Averaging

(BMA) provides the existence of an N-shaped pattern between national income and environmental pollution, and GDP plays a crucial role in carbon dioxide ($CO_2$) emissions (Mitsis 2021). It has been discerned that each year almost 0.8 million deaths and about 4.6 million people have lost their lives due to air pollution (WHO 2016). $PM$, the mass concentration of all aerosol particles smaller than 10 $\mu m$ in diameter is effectively susceptible to penetration into the alveolar region (Harrison 2000). It is a well-established fact that $PM$ concentrations are possibly attached to different types of life-threatening viruses and bacteria, which breaks down the human immune system entirely (Zoran 2020). Thus, to alleviate pollution levels, controlling the $PM$ concentration is an effective initial step to improve the air quality index (AQI) (Kumar et al. 2015). After the implementation of compressed natural gas (CNG) fuel, carbon monoxide ($CO$), and sulfur dioxide ($SO_2$) decreased, but there was an increase in $PM_{10}$, nitrogen oxide ($NO_x$) levels (Ravindra 2006). For high traffic locations, the level of $PM_{2.5}$ has run over the standard $PM_{2.5}$ level, especially in winter (Pant 2015), and the predicted number of road vehicles will reach 25.6 million by 2030 (Kumar et al. 2011). Some studies found that pollution affects children because of their weak immunity system (Schwartz 2004) and increases the risk of gestational diabetes mellitus (GDM) in pregnant women (Robledo 2015). Pollution at an alarming level attracts researchers' recognition so much so that the influence of different policies, for example, odd and even trials, are reviewed in detail, and are considered effective in mitigating pollution during the early morning and late evening hours (Kumar et al. 2017)

From previous studies, we are aware of the deadlier impacts of air pollution and the necessary measurements to control these pollutant concentrations in the air. However, to mitigate pollution, different policies will be effective when we predict the trend of pollutant concentrations in the air with better accuracy. Only then can proper measures be taken to control this pollution. Therefore, a review of the previously used methodologies is very important. $SO_2$ and total suspended particles (TSP) were effectively correlated with temperature, wind speed, atmospheric pressure, and relative humidity during the winter season from 1999 to 2005 (Ilten 2008). In Turkey, the AQI was improved by banning use of hard coal of poor quality for domestic purposes, and the $SO_2$ and TSP concentrations decreased in the air with increasing temperature, pressure, and wind speed. Time series forecasting and data mining techniques have been implemented to show the patterns of several pollutant concentrations (Sharma et al. 2018). In that study, it was indicated that in the future, a heavy load in transportation would be a key factor contributing to significant $NO_x$ emissions and crop burning, and construction work would be mainly responsible for high concentrations of $PM_{10}$ and $PM_{2.5}$ concentration in the air. An exhaustive statistical analysis of pollution data is also available; for example, seasonal influence in daily time data, at what time the pollution is at its peak, is justified statistically (San Martini et al. 2015). In that study, it was detected that in Beijing, the pollution is minimum in spring, and in the remaining cities, the pollution is minimum during the summer. The pollutant concentration is at its peak in Beijing during the night, whereas in the remaining cities, the pollution is generally maximum during rush hours in the morning (San Martini et al. 2015). Combining geospatial interpolation methods with ML models, researchers have effectively predicted mud content in the southwest part of Australia. For this purpose, they used RFK and random forest inverse distance square (RFIDS), and detected that these hybrid models are capable of interpolating the measurement of accuracy because they reduce RMSE by 30% and 19%, respectively (Li 2011). Researchers have combined bidirectional long short-term memory (BLSTM) networks with IDW to exhibit a new methodology, that is, IDW-BLSTM. This helps analyse the long-term temporal upshot of pollution (Ma 2019). Many researchers have attempted to develop new models for acquiring authenticity. There are three main types of accessible models: (i) deterministic models, for example, the operational street pollution model (OSPM) (Berkowicz 2000), (ii) statistical models, namely the multiple regression model (MLR) (Berkowicz 2000), and (iii) machine learning models, such as random forest (RF) (Hengl 2007). In Table (1) there is a precise and foremost recapitulation of previous research works regarding the methodologies used in the previous case study.

Table 1: Important methodologies in previous research work.

| Location | Focus area | Prime detection | Reference |
|---|---|---|---|
| Delhi | Estimation of $PM_{2.5}$. | IDW and OK | (Shukla 2020) |
| Guangdong province, China | Spatio-temporal estimation of $PM_{2.5}$. | IDW-BLSTM performed efficiently | (Ma 2019) |
| China | Spatio-temporal estimation of $PM_{2.5}$ | RFSTK responded satisfactorily | (Shao 2020) |
| China | Spatial and temporal pattern recognition of $PM_{2.5}$ concentration | Linear regression and grey system correlation analysis | (Lu 2017) |
| Delhi | Relationship between $PM_{2.5}$ and other spatio-temporal covariates are explained | Six ML learners | (Mandal 2020) |
| China | Statistical analysis of $PM_{2.5}$ concentration | Seasonal influence | (San Martini 2015) |

In the year, 2020 the situation was different from that in previous years. The COVID-19 pandemic, caused many countries to institute social lockdown measures to break the chain of this virus. In India, the central government announced a lockdown with the same intention. The social and transportation movement was fully restricted, except in some emergencies because of the "Janata Curfew" announced by the honourable Indian Prime Minister. After announcing the lockdown, it was observed that the pollution declined on a large scale. Many researchers have detected a relationship between COVID-19 and air pollution. They proposed lockdown as a measure of controlling pollutant emissions because air pollution significantly contributes to weakening the human immune system, causes respiratory hazards, and creates a greenhouse effect. The use of interrupted time series modelling to test the significant influence of lockdown in changing the pollutant concentration has been discussed (Cameletti 2020). The lockdown intervention reduced nitrogen-di-oxide ($NO_2$) and $PM_{10}$ concentration in the city of

Brescia significantly (Cameletti, 2020). In Wuhan, a drastically decreasing tendency is noticed in pollutant concentrations, particularly $NO_2$ and $SO_2$. Just prior to the lockdown, the $NO_2$ concentration was $23.1 - 40.7$ $\mu g/m^3$, whereas after lockdown, that concentration varied between $13.8\text{-}28.6$ $\mu g/m^3$ due to the strict restriction of transportation movement (Brimblecombe 2020). Bashir (2020) found that average temperature, minimum temperature, and air quality are linearly related to the COVID-19 lockdown. It has stopped the spread of the COVID-19 virus in New York, USA, and has improved air quality. The improved air quality indicates that green environmental policies should be promoted because they mitigate the spread of infectious diseases, such as COVID-19. Shrestha (2020) observed a significant decline in $PM_{2.5}$, $PM_{10}$, and $NO_2$ concentrations in cities such as Bangalore, Beijing, Bern, Delhi, Lima, London, Madrid, New York, Paris, Seoul, Sydney, Tokyo, Ulaanbaatar, and Vienna. The National Air Quality Index (NAQI) was notably improved during the lockdown (DL) period in megacity Delhi, and the average $PM_{10}$ and $PM_{2.5}$ concentrations in the air decreased by 57% and 33%, respectively, on average, compared to the previous three years (Mahato 2020). Moreover, after the commencement of the first day of lockdown, the air quality improved by 40%, and approximately 54%, 49%, 43%, 37%, and 31% decline in AQI was seen in the central, eastern, southern, western, and northern regions of NCT Delhi, respectively (Mahato 2020). $PM_{2.5}$ saw the largest reduction among the other important pollutant concentrations in most of the cities in India, and AQI was lower compared to the previous years in India (Sharma et al 2020). The AQI improved by approximately 30%– 46.67%, and there was a huge reduction in the $PM$ concentration level in Delhi, followed by UP and Haryana, because of fewer vehicles on the road and reduced industrial emissions (Gautam 2021). The researchers evaluated spatio-temporal variations of pollutants in Delhi, Mumbai, Chennai, Bangalore, and Kolkata over four time periods: March 2019 –April 2019 and the same period in 2020. The other time points were 10th March 2020 to 20th March 2020, considered as before lockdown (BL) and 25th March 2020 to 6th April 2020 (DL), which highlighted a statistically significant decline in all the pollutants (Jain 2020).

Although the proposed temporal trend and spatial interpolation models performed well, they had some limitations. Previous research has ignored whether the relative temporal change of pollutant concentrations in the air of a particular time window depends only on the window size or time point. If it depends on time, it is obvious that seasonal influence and secular trend will exist. If it is independent of time, measuring the seasonal effect of pollutant concentration does not make any sense. In addition, when researchers analyse the trend of pollutant levels in the air, they neglect to explain the pollutant concentration trend. Let us assume that in winter, the trend of $PM_{2.5}$ is increasing (Mahato 2020), but the question is whether crop burning during winter is the sole cause of this trend, or there are some other issues which are also responsible for this trend. Likewise, before applying the spatial interpolation model, there is no information to validate whether the covariates vary spatially, because if the datasets are not spatially correlated, then it is easy to interpret that the monitoring stations are located far from each other. As a result, we miss the spatial information contained in the neighbourhood of each monitoring station. Similarly, when they measured the spatio-temporal impact of lockdown, they compared the averages of pollutant concentrations and concluded about the spatio-temporal effect of lockdown. However, when comparing the averages of the pollutant concentrations, they include the intervention of lockdown and seasonal land use information simultaneously, which is an ambiguous measurement. Limited information is available regarding the frequency distribution of pollutants. As a result, the probabilistic distributions are fully ignored in the context of air pollution in Delhi, and the statistical discussion of the effectiveness of the major activities and lockdown in $PM_{2.5}$ emission is omitted. In addition, at the time of fitting the variogram models, there was no concrete discussion of Delhi air pollution.

Therefore, to mitigate the above shortcomings, we first discuss the features of the frequency curves of daily $PM_{2.5}$ emission. Thereafter, we discuss the effectiveness of lockdown and the importance of major activity at $PM_{2.5}$ level with statistical logic (Figure 1).

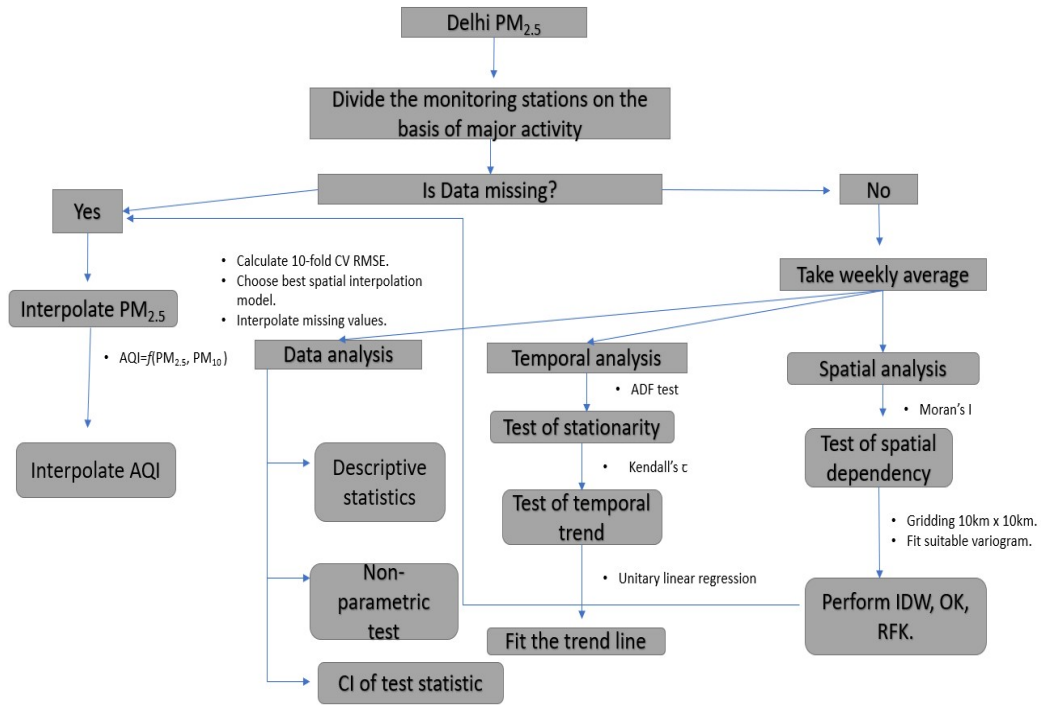

Figure 1: Detail of the research work.

Before conducting the temporal trend analysis, we checked the stationarity of the time-series data. We then studied the temporal trend and seasonal influence of the dataset. We discussed the data for which weeks are spatially correlated using Moran's $I$ index. Subsequently, we interpolated the $PM_{2.5}$ value of unobserved locations using three popular spatial interpolation techniques, IDW, OK, and RFK, using 10 km × 10 km spatial resolution, and compared them based on the RMSE. We estimated the lock-down effect in major activity zones using the (DID) estimator. Thereafter, we emphasise the spatio-temporal behaviour of each monitoring station for each season.

## 2. Data and Methodologies

### 2.1 Study Area

Here, we selected Delhi, the capital of India, to study the air pollution of BL and DL during the first wave of the COVID-19 pandemic in India. Being the capital of India, rapid urbanisation, increasing amounts of traffic, increasing population, and energy consumption at an alarming level are mainly responsible for air pollution in Delhi. Sometimes, the $PM_{2.5}$ concentration in the air reaches 999 $\mu g/m^3$ (Mukherjee 2018). A spike in the vehicle count in Delhi has been identified as the cause of higher pollutant concentrations in the air (Samal 2013). Among all air pollutants, $PM_{2.5}$ especially affects public health (Zheng 2015).

### 2.2 Site Selection

We considered the air pollution data collected by the monitoring stations, maintained by the Central Pollution Control Board (CPCB), Delhi Pollution Control Committee (DPCC), and the Indian Institute of Tropical Meteorology (IITM). For our research purpose, we collected data on several air pollutants, such as $PM_{2.5}$, $PM_{10}$, $CO$, and $NO$, from the CPCB websites. To map the spatio-temporal distribution of air quality in Delhi and to contrast between the $PM_{2.5}$ concentrations of the BL and DL periods, these data play a key role. In this dataset, there are 38 monitoring stations, which are displayed in Figure (2), where the data were collected over 24 h.

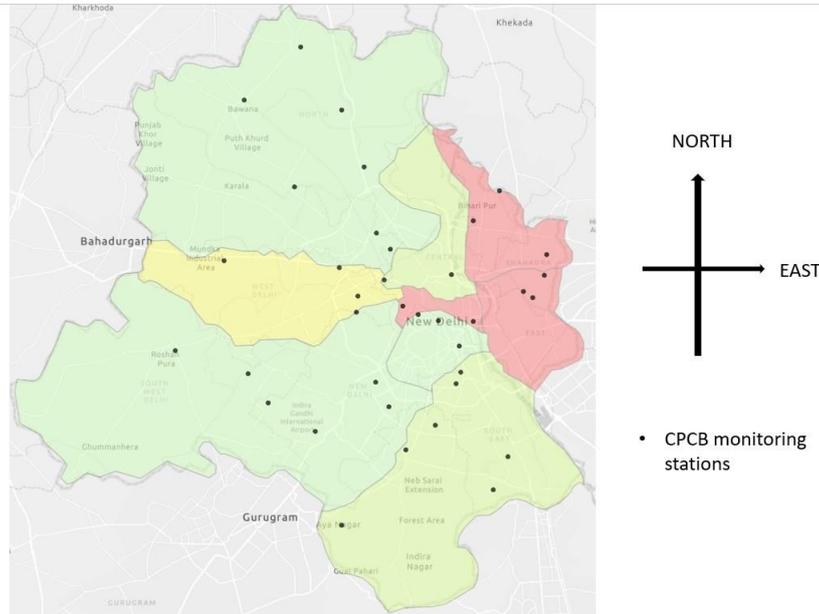

Figure 2: List of monitoring stations in Delhi

In this study, the time period was taken from $1^{st}$ January 2019 to $28^{th}$ February 2021 as a result of which the spatio-temporal impact of lockdown on $PM_{2.5}$ is visible. However, there are two shortcomings in applying spatial interpolation techniques to estimate $PM_{2.5}$ in Delhi NCR because of the unavailability of data and the distance between Delhi, NCR, and other monitoring stations. Among the 38 monitoring stations, there is no data available for $PM_{2.5}$ in East Arjun Nagar, Delhi, and the data for the remaining stations are sometimes missing, such as in Anand Vihar from $4^{th}$ April 2020 to $18^{th}$ April 2020 and from $1^{st}$ May 2020 to $14^{th}$ May 2020 in DTU, the data are missing from $3^{rd}$ February 2019 to $6^{th}$ February 2019 and from $28^{th}$ March 2020 to $31^{st}$ March 2020. Based on the spatial analysis, some of the stations are divided based on major acts, namely, transport, residential, commercial, and institutional zones, as shown in Table (2). In Table (2), we consider Anand Vihar, IGI Airport, and ITO Delhi as the major transport zones; R K Puram, Punjabi Bagh, Ayanagar, Burari Crossing, and Sirifort as major residential zones; Mandirmarg, Lodhi Road, and Shadipur as major commercial zones; and lastly CRRI, Mathura Road, DTU, NSIT Dwarka, North Campus DU, and Pusa as major institutional zones.

Table 2: Major activity of monitoring stations.

| S.No. | Monitoring Station | Major Activity |
|---|---|---|
| 1 | AnandVihar | Transport |
| 2 | IGI Airport, T3 | Transport |
| 3 | ITO, Delhi | Transport |
| 4 | R K Puram | Residential |

| 5 | Punjabi Bagh | Residential |
| 6 | Ayanagr | Residential |
| 7 | Burari Crossing | Residential |
| 8 | Sirifort | Residential |
| 9 | Mandirmarg | Commercial |
| 10 | Lodhi Road | Commercial |
| 11 | Shadipur | Commercial |
| 12 | CRRI, Mathura Road | Institutional |
| 13 | DTU | Institutional |
| 14 | NSIT, Dwaraka | Institutional |
| 15 | North Campus DU | Institutional |
| 16 | Pusa | Institutional |

## 2.3 Spatio-Temporal Methods

In this study, we examine the characteristics of $PM_{2.5}$ emission in the air, and we measure the statistical significance of lockdown on air pollution. In this section, we present a brief overview of the statistical methods for spatial, temporal, and spatio-temporal impact analyses.

Let $y_{ts}$ be the observed value of the dependent variable at the $t^{th}$ time point at the $s^{th}$ location, where $t = 1,2,3,\ldots,n$ and $s = 1,2,3,\ldots,d$. Likewise, $x_{ts}^{(i)}$ be the observed value of the $i^{th}$ independent variable at the $t^{th}$ time point on $s^{th}$ location. We also studied the stationarity of the data. If the ACF of the data converges to 0, then the data are strictly stationary; otherwise, they are non-stationary (Brockwell, 2002). If the data are non-stationary, then trends, seasonality, cyclical fluctuations, and randomness exist in the data. Let us assume that the temporal cross-section of the dataset is additive in time, and the additive time series model is

$$y_{ts} = T_{ts} + S_{ts} + C_{ts} + R_{ts} \qquad (1)$$

In Equation ((1)),

$T_{ts}$ = Measurement of the secular trend at $s^{th}$ location at $t^{th}$ time point.
$S_{ts}$ = Measurement of seasonality at $s^{th}$ location at $t^{th}$ time point.
$C_{ts}$ = Measurement of cyclical fluctuation at $s^{th}$ location at $t^{th}$ time point.
$R_{ts}$ = Measurement of irregularity at $s^{th}$ location at $t^{th}$ time point.

Then, we measured the seasonal influence of $t^{th}$ time point on the $y_{ts}$ using the ratio-to-moving average method. We analysed the temporal trend of the $y_{ts}$ for $t^{th}$ time point for every geographical location during the two years.

After analysing the temporal behaviour, we focused on the spatial trend of $y_{ts}$. In this situation, we perform the test of spatial auto-correlation using Moran's I index at 0.05 level of significance (Till 2018) where

- $H_0$: The data are not spatially autocorrelated ($I = 0$).
- $H_1$: The data are spatially autocorrelated ($I \neq 0$).

If the data are spatially auto-correlated, then we perform a spatial trend analysis of the $y_{ts}$. In this scenario, the main challenges are: (1) The unobserved gridded data points (10 km × 10 km) and (2) The missing values. Therefore, we first concentrated on the spatial interpolation. For this purpose, we used three well-known geospatial interpolation techniques and compared their interpolation accuracies. These three spatial interpolation methods are (i) IDW, (ii) OK, and (iii) RFK. In IDW (Wackernagel, 2003), the interpolated $y_{ti}^*$ is the unobserved $y_{ts}$ where $s = i$. This is the unobserved value at the $i^{th}$ location at the $t^{th}$ time point. The IDW equation can then be expressed as follows:

$$y_{ti}^* = \frac{\sum_{s \in N(i)} w_s^{IDW} y_{ts}}{\sum_{s \in N(i)} w_s^{IDW}} \quad (2)$$

In Equation ((2)), $N(i)$ is the neighbourhood point of the $i^{th}$ location, and $w_s^{IDW} = \frac{1}{d(s,i)^p}; s \neq i$ and $s \in N(i)$. In Equation ((2)) the performance of IDW is dependent on the choice of $p$. Similarly, in OK, (Cressie 2015) the equation can be written as:

$$y_{ti}^* = \sum_{s=1}^{d} w_s^{OK} y_{ts} \quad (3)$$

In Equation ((3)), $w_s^{OK}$ is calculated using the variance-covariance function fitting an appropriate variogram with better accuracy, and this variogram is chosen based on the minimum RMSE. These two methods might look identical, but by their origin, they are completely different, as IDW is purely the deterministic approach, whereas OK is purely the probabilistic approach. Now we think about a new hybrid geospatial interpolation technique that is efficient enough regarding the sense of interpolation, that is, RFK.

RFK combines two methods: $(i)$ RF to fit the explanatory variables, and $(ii)$ OK to fit the OOB errors with expectation 0 (Hengl 2007). Let $\vec{x}_0$ be the unobserved location and $\vec{x}_i$ be the $i^{th}$ neighbourhood location of $\vec{x}_0$ where $i = 1, 2, \ldots, n$ then the RFK (Shao 2020) model is:

$$y_{ti}^* = \sum_{s \in N(i)} \sum_{j=1}^{m} \alpha_j x_{ts}^{(j)} + \sum_{s \in N(i)} w_s^{RFK} \epsilon_{ts} \quad (4)$$

where, $\alpha_j$ is the RF regression coefficient, $w_s^{RFK}$ is the kriging weights and $\epsilon_{ts}$ are the OOB errors of the $s^{th}$ location. Using the bootstrapped sample, RF performs regression between the explanatory variables by building a huge collection of trees randomly which do not correlate (Breiman 2001). It provides important measures such as the Gini-Mean decrease (G) (Grömping, 2009), and error rate. Thereafter, we considered the spatial features of the OOB error and interpolated the OOB errors using the OK. The RF model is fitted on the data in R using the "randomForest" package (RColorBrewer 2018). In this function, there are two important parameters: ntree and mtry, where ntree denotes number of uncorrelated decision trees and mtry denotes number of the independent splitting variables. The explanatory variables are listed in the following Table (3).

Table 3: List of variables used in RFK model and their corresponding measuring units.

| Symbol | Variable details | Unit |
|---|---|---|
| NO | Nitric Oxide | $\mu g/m^3$ |
| $NO_2$ | Nitrogen Di-Oxide | $\mu g/m^3$ |
| $NO_x$ | Other Nitrogen Oxides | $\mu g/m^3$ |
| CO | Carbon Mono Oxide | $\mu g/m^3$ |
| $PM_{10}$ | Particulate Matters with a | $\mu g/m^3$ |
| | diameter of 10 micrometers | |
| $PM_{2.5}$ | Particulate Matters with a diameter of 2.5 micrometers | $\mu g/m^3$ |
| East | Easting Position longitude | $/\ ^0N$ |
| North | Northing Position latitude | $/\ ^0E$ |

We now consider the spatio-temporal intervention of lockdown on $PM_{2.5}$ emission. Therefore, we used the DID regression (Mascha 2019) technique to estimate the treatment effect by comparing the differences in the outcomes of the pre-and post-intervention periods and the outcomes between the intervention and control groups (in this case, the treatment group was lockdown and the control group was the major activity zone). DID regression can be expressed as follows:

$$E(y_{ts}|Z_s, I_t) = \beta_0 + \beta_1 \cdot I_t + \beta_2 \cdot Z_s + \beta_3(I_t \cdot Z_s) + \epsilon_{ts} \quad (5)$$

In the above Equation ((5)), $Z_s$ is the control group, $I_t$ is the indicator variable which is equal to 1 if the treatment group is present and 0 if the treatment group is absent, $\beta_i; i = 0,1,2,3$ are the linear regression coefficients and $\epsilon_{ts}$ are the residuals of this linear model. $E(y_{ts}|Z_s, I_t)$ is the conditional expectation of $y_{ts}$ given $Z_s$ and $I_t$ which measures the conditional average of $y_{ts}$ in the presence of $Z_s, I_t$. In Equation ((5)), $\beta_3$ measures the interaction effect of $Z_s$ and $I_t$ on $y_{ts}$. In this study, $\beta_3$ estimates the spatio-temporal effect of the lockdown of $PM_{2.5}$ in the major activity zones.

2.4 Model Accuracy

The accuracy of the models are validated by the following three methods :

1. $R^2$, that is, the coefficient of determination

2. RMSE

3. K-fold CV

Suppose that $y_{ts}$, $\hat{y}_{ts}$ and $\bar{y}$ are the observed and predicted values of the variable at the $t^{th}$ time point on the $s^{th}$ location and the mean of the values of the variable, respectively. The data are available for $d$ locations and $n$ time points. Then, the sum of squares of errors ($SSE$), mean square error ($MSE$), and total sum of squares ($TSS$) are defined as follows:

$$SSE = \sum_{t=1}^{n} \sum_{s=1}^{d} (y_{ts} - \hat{y}_{ts})^2 \quad (6)$$

$$MSE = \frac{\sum_{t=1}^{n} \sum_{s=1}^{d} (y_{ts} - \hat{y}_{ts})^2}{nd} \quad (7)$$

$$TSS = \sum_{t=1}^{n} \sum_{s=1}^{d} (y_{ts} - \bar{y})^2 \quad (8)$$

Then, using Equation ((6)), ((8)), and ((7)), the measurements of accuracy are respectively $R^2 = 1 - \frac{SSE}{TSS}$ and RMSE=$\sqrt{MSE}$, where $R^2$ denotes what proportion of $PM_{2.5}$ is explained by the model, and in contrast, RMSE denotes the unexplained variability. Here, to find the value of $p$ in IDW, $R^2$ and RMSE play a prime role (minimum RMSE and maximum $R^2$) and to validate the K-fold CV, K was taken as 10. The 10-fold CV indicates that the dataset is divided into 10 data sets at random, and among these datasets, 9 data sets were taken as the training dataset, and 1 data set was taken as a test dataset. Likewise, to apply the variogram model, we have determined salient parameters such as range, nugget, and sill aiming towards the minimum RMSE. To uphold the OK predicted result, we brought about a 10-fold CV. In RFK, after interpolating the OOB residuals, the final predicted value was ratified by the RMSE.

## 3. Result and Discussion

### 3.1 Descriptive Studies

This study included 25,875 data points in this research work. We have noticed a descriptive summary of the data in Table (4) in a nutshell. During the entire period, the minimum and maximum values of $PM_{2.5}$ during the entire period of time are respectively 10.22 $\mu g/m^3$ and 715.04 $\mu g/m^3$. As a result, we can understand that there exists a huge variation in 24 hour $PM_{2.5}$ values in the air of Delhi. To capture the entire dispersion of the data, we studied the standard deviation of $PM_{2.5}$, which was $75.18 \mu g/m^3$. In the entire data set, the first 25 percentile of daily $PM_{2.5}$ data are lying under 41.88 $\mu g/m^3$, the 50 percentile of daily $PM_{2.5}$ data are under 67.77 $\mu g/m^3$, and the 75 percentile of daily $PM_{2.5}$ data are under 122.14 $\mu g/m^3$. In addition, the mean and mode of the daily $PM_{2.5}$ data are respectively 94.12 $\mu g/m^3$ and 45.044 $\mu g/m^3$ (from Table (4)). For the entire dataset, the mean was greater than the median, and the median was greater than the mode. Pearson's first measure of skewness is 0.6527736, which indicates that the frequency curve of $PM_{2.5}$ is positively skewed, which is the degree of departure from the symmetry. Therefore, the frequency curve had a higher density in the smaller values of 24-hour $PM_{2.5}$ emission data than the larger values of 24-hour $PM_{2.5}$ emission data, and after the mode of the data, the density of the frequency curve decreased with increasing values of $PM_{2.5}$. The kurtosis of $PM_{2.5}$ is 2.211476, which is greater than 0. Therefore, the distribution of $PM_{2.5}$ is leptokurtic, indicating that there is a high density in the neighbourhood of the daily $PM_{2.5}$ data (i.e. 45.044 $\mu g/m^3$ from Table (4)).

Table 4: Descriptive statistical measurements of 24 hour $PM_{2.5}$ emissions during the entire time period.

| Summary | Value ($\mu g/m^3$) |
| --- | --- |
| Minimum | 10.22 |
| First Quartile ($Q_1$) | 41.88 |
| Median ($Q_2$) | 67.77 |
| Mean | 94.12 |
| Third Quartile ($Q_3$) | 122.14 |
| Maximum | 715.04 |
| Mode | 45.044 |
| Standard Deviation | 75.18 |

For better visualisation and proper understanding, the purpose histogram of $PM_{2.5}$ of the four major monitoring stations is shown in Figure (3), which is sufficient to show that the graph is positively skewed. In this Figure (3), we just show the pattern of the data of some sample monitoring stations to get a brief idea about the pattern of daily $PM_{2.5}$ emissions in each monitoring station. In this figure, we plotted the basic histogram of Anand Vihar, R K Puram, Mandir Marg, and CRRI Mathura. In this Figure (3), we find that the minimum daily $PM_{2.5}$ emission is being detected in R K Puram (i.e. 11.38 $\mu g/m^3$), and maximum daily $PM_{2.5}$ emission is being identified in Anand Vihar (i.e. 373.46 $\mu g/m^3$). In Anand Vihar, the maximum variation is detected because Anand Vihar is one of the paramount transport zones (Figure (3)) with heavy flow of vehicles and the daily $PM_{2.5}$ concentration in the air is very high. Similarly, because of the COVID-19 lockdown, transportation was restricted; as a result, there was a massive decline in 24 hour $PM_{2.5}$ emission. Therefore, owing to the large variation, the standard deviation was significant for Anand Vihar.

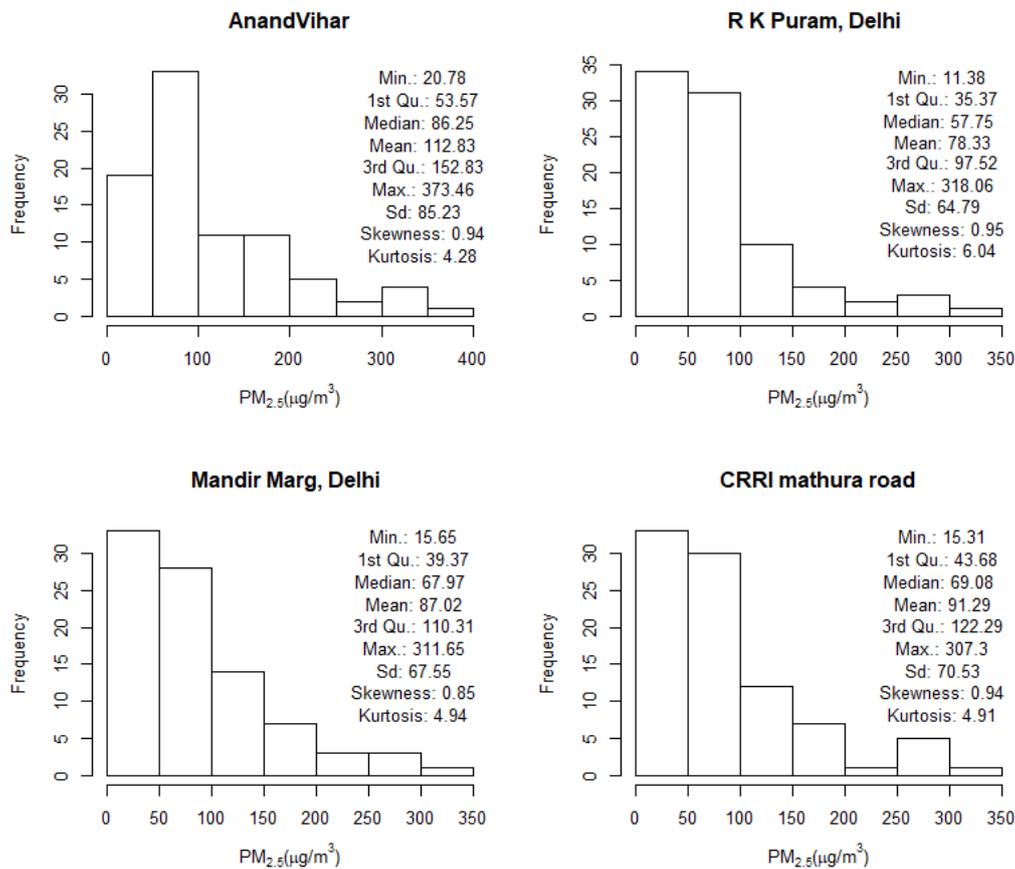

Figure 3: Histogram of PM $_{2.5}$ of monitoring stations.

The CI of mean $PM_{2.5}$ during the cumulative time span was 91.37-96.87 at a confidence level of 0.95. Thus, it can be concluded that people are at a high risk of exposure to unhealthy air. In this study, we divided the entire time period into the following three parts in Table (5).

Table 5: Division of entire time periods

| Time Period | Starting Date | Starting Week | Ending Date | Ending Week |
|---|---|---|---|---|
| Before Lockdown (BL) | 17-03-2019 | 12 | 29-06-2019 | 26 |
| During Lockdown (DL) | 22-03-2020 | 12 | 27-06-2020 | 26 |
| After Lockdown (AL) | 28-06-2020 | 27 | 29-08-2020 | 36 |

Then, we calculated the weekly average of $PM_{2.5}$ for each week and compared the weekly averages of $PM_{2.5}$ values. The CI of the mean difference of BL, DL; BL, AL; and DL, AL are respectively 22.86-28.16, 44.68-49.87, and 19.57-23.96 at a significance level of 0.05. This result is very useful for distinguishing the pattern of $PM_{2.5}$ emissions during the three time periods. This study leads us to one of the important remarks that the features of $PM_{2.5}$ emission during BL and AL are significantly different from each other because it is the CI of the mean difference of the weekly average of $PM_{2.5}$ emission for the pre-and post-lockdown time periods. The results showed that the lockdown had an influential effect on $PM_{2.5}$ emission. Therefore, after lockdown was over, the $PM_{2.5}$ level in the air was less than that for the BL time period.

3.2 Relationship between the variables

In this section, we determine the linear relationship between the different types of pollutants so that during the application of RFK, we can skillfully select the important covariates to predict $PM_{2.5}$ emission. In Figure (4), we detect that the pollutants in the air are highly correlated, and we assume $r$ as the measure of correlation. We observe that the daily $PM_{2.5}$ emission and daily $PM_{10}$ emission during the entire time period are positively correlated, that is, 0.98 which indicates that if $PM_{2.5}$ increases, then $PM_{10}$ will also increase. Similarly $NO_2$, $NO_x$ are also positively correlated, with a correlation coefficient of 0.91.

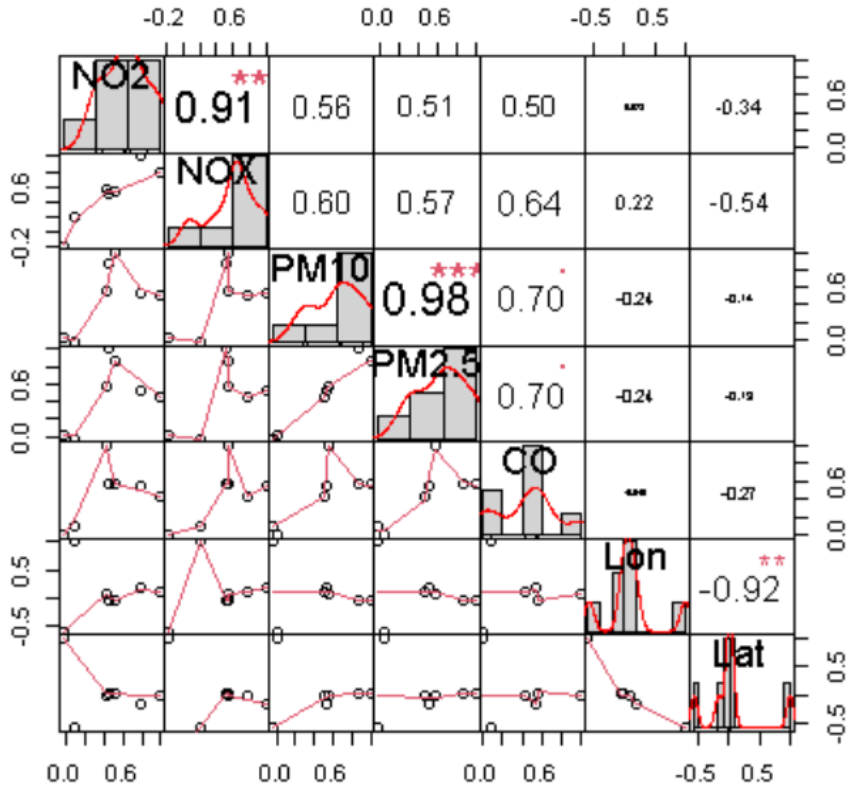

Figure 4: Correlation Matrix of the variables to detect the degree of association between them.

**Note:** " *** " for "$p < 0.001$", " ** " for "$p < 0.01$", " * " for " $p < 0.05$ ", and "." for " $p < 0.1$".

Similarly, the daily emissions of $PM_{10}$ and $PM_{2.5}$ are also positively correlated with CO, that is , 0.70. Apart from these, $NO_2$ is positively correlated with $PM_{10}$ with a correlation coefficient of 0.56. Similarly, $NO_2$ is positively correlated with $PM_{2.5}$ and CO. The correlation coefficient between $NO_2$ and $PM_{2.5}$ is 0.51 and that between $NO_2$ and CO is 0.50. Another important pollutant $NO_x$ is highly correlated with $PM_{2.5}$, with a correlation coefficient of 0.57, and is also positively correlated with $PM_{10}$ with a correlation coefficient of 0.60. The correlation coefficient between $NO_x$ and CO was 0.64. There is no significant correlation between the explanatory variables and the latitude and longitudinal position; however, the missing correlation between the pollution concentration and the values of the latitudes or longitudes does not guarantee the absence of influence in the case of nearby stations. As a result, any monitoring station can influence the daily pollutant concentrations at another nearby monitoring station. Therefore, we were able to take enough gridded points between two monitoring stations to capture the entire spatial information, which will help interpolate the pollutant concentration on the unobserved locations.

4. Temporal Trend Analysis

It is important to analyse whether the data are stationary before opting for a temporal trend analysis. The temporal stationarity of the dataset implies that the temporal association of the variable of a time window is dependent only on the window size. If it is non-stationary, then the association of a time window depends not only on the window size but also on the starting and ending time points. For this reason, the autocorrelation function (ACF) plays a major role if it does not converge toward 0 while the lag or window size is increasing; then, it can be surmised that the time series data is not stationary. Generally, in the stationary dataset, the ACF converges to 0 when the lag is increased. In Figure (5), we study the variation in ACF with respect to the lag
starting from 0 to 25.

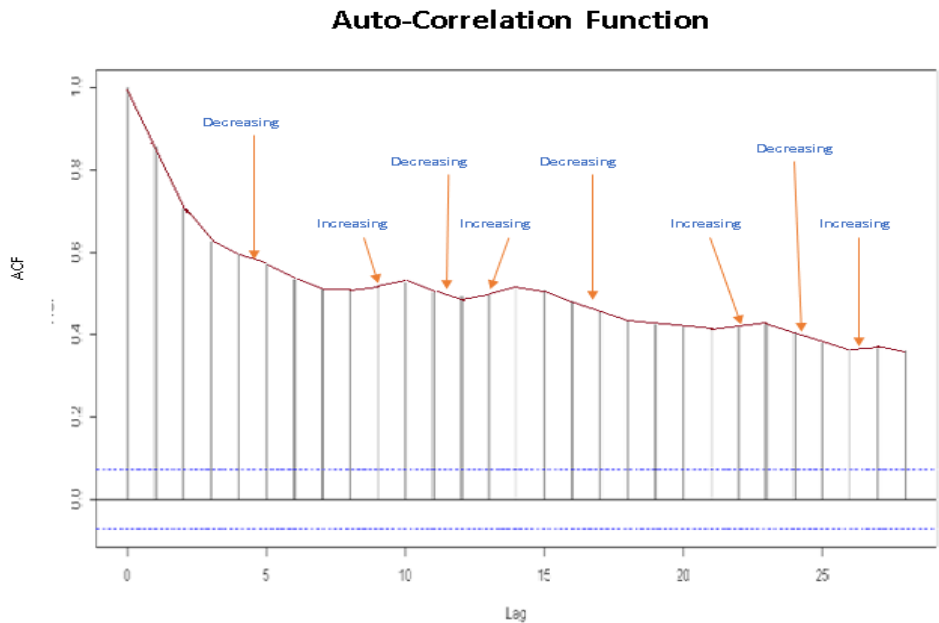

Figure 5: ACF of the entire data.

We notice that the ACF decreases and increases alternatively from Figure (5) with respect to lag, and the ACF does not go to 0 while the lag is increasing. Therefore, we can infer that the data are not stationary. It is possible to conclude that the distribution of $PM_{2.5}$ is dependent on the starting and ending time points, and there is a significant correlation of $PM_{2.5}$ values for each day.

As the data are not stationary, seasonality may exist. We now discuss the seasonal influence in each of the major activity zones. For this purpose, we consider February and March as spring; April, May, June, and July as summer; August and September as monsoon; and October, November, December, and January as winter. In Table (6), we detect that during winter in every zone, the seasonal effect is very high compared to the other seasons. For example, the effects of winter during 2019 on the major activity zones such as transport, residential, commercial, and institutional zones were 60.40, 51.52, 48.07, 56.18 respectively. The values for 2020 are 62.36, 53.78, 50.23, and 58.14, respectively. Likewise, the seasonal influence of monsoon on the transport, residential, commercial, and institutional zones in 2019 were -68.34, -57.44, -55.09, -65.42, and similarly, and those during 2020 were -68.44, -57.27, -55.17, and -65.61, respectively. The positive value of seasonal influence (from Table (6)) indicates that there is a positive seasonal trend, and the negative value of seasonal influence indicates that there is a negative seasonal effect on $PM_{2.5}$. For example, in the transport zone during monsoon in the year 2020, the seasonal influence was −68.44, which indicates a decreasing seasonal effect, but during winter, it was 62.36 indicating a positive seasonal impact on daily $PM_{2.5}$ emission. However, owing to the COVID-19 lockdown and restrictions on public transportation, seasonal influence is reduced. Therefore, in the transport zone during spring in the year 2019, the seasonal influence was -2.29, whereas during spring in the year 2020, it was considerably less than the previous one, that is, -3.21. The same feature is also observed in the remaining zones (Table (6)). Along with these, we see that every year during winter, average $PM_{2.5}$ value is comparatively higher than that during the monsoon. For example, in the institutional zone, the average $PM_{2.5}$ values reach up to $190.71 \mu g/m^3$ during winter, whereas during monsoon, this reaches $41.56 \mu g/m^3$.

Table 6: Seasonal Influence of each season on daily $PM_{2.5}$ emission in four major activity zones.

| Season | Year | Transport Zone | Residential Zone | Commercial Zone | Institutional Zone |
|---|---|---|---|---|---|
| Summer | 2019 | -54.92 | -47.32 | -42.70 | -52.78 |
| Monsoon | 2019 | -68.34 | -57.44 | -55.09 | -65.42 |
| Winter | 2019 | 60.40 | 51.52 | 48.07 | 56.18 |
| Spring | 2019 | -2.29 | -2.12 | -2.65 | 2.38 |
| Summer | 2020 | -54.71 | -47.08 | -42.48 | -52.09 |
| Monsoon | 2020 | -68.44 | -57.27 | -55.17 | -65.61 |
| Winter | 2020 | 62.36 | 53.78 | 50.23 | 58.14 |
| Spring | 2020 | -3.21 | -2.94 | -3.48 | 0.742 |

The temporal pattern analysis of the four zones of the daily average of $PM_{2.5}$ values for three described time periods, namely, pre-lockdown, during-lockdown, and post-lockdown, is shown in Figure (6).

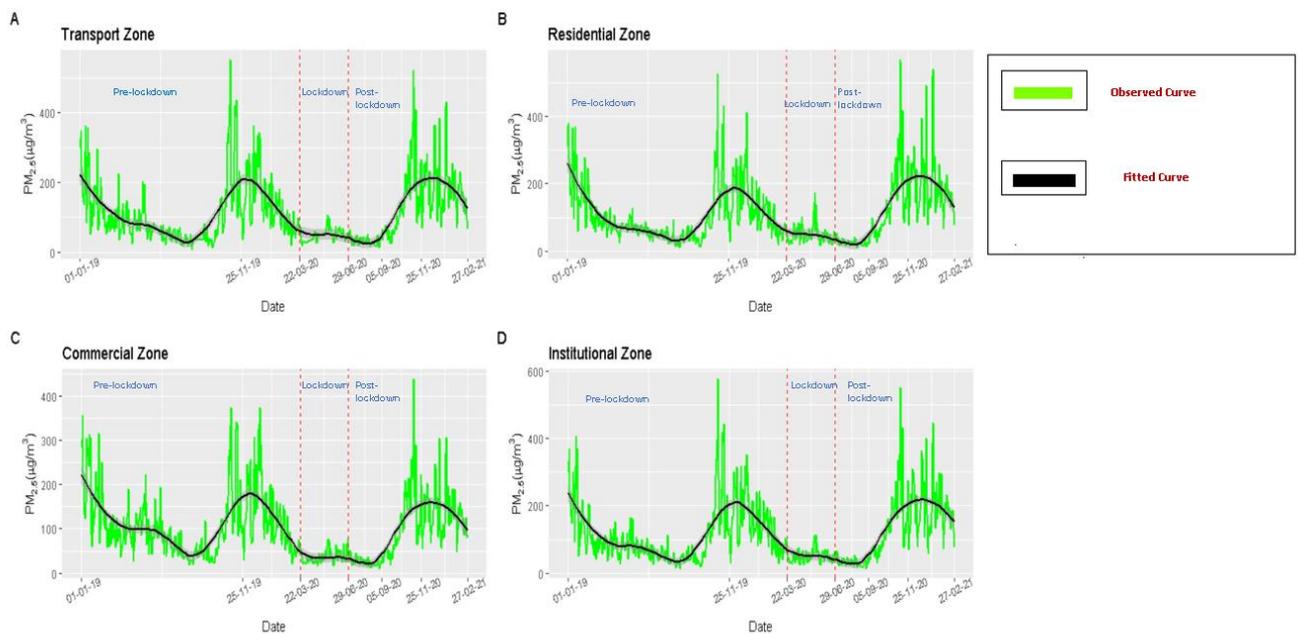

Figure 6: Time series trend for four zones during the entire time period.

In Figure (6), we can see that in every major activity zone, the trend is decreasing with respect to the time up to the end of monsoon. After the monsoon season, it starts to increase and reaches its maximum, especially at the end of November. However, during the lockdown period, the daily $PM_{2.5}$ value was comparatively lower than that in the summer of the previous year. However, in the post-lockdown time period, the trend started to increase slowly after the monsoon, and it reached the peak in mid-November; however, given the social and transport restrictions, this peak is smaller than the peak point of the year 2019, which is considered pre-lockdown time period, and after winter, it began to decrease. As a result, we can infer (from Figure (6)) that, along with the seasonal impact, secular trend, cyclical, and random effects, one more effect is being confounded: the lockdown effect which will be estimated in the later part of this research.

We now compare the average amount of important air quality parameters (such as $PM_{2.5}$, $PM_{10}$, and AQI) for each of the major activity zones before and during the lockdown period. In this situation, we calculated the average pollutant concentrations and AQI levels for the BL and DL time periods, which are mentioned earlier in Table (5). Then, we subtracted the pre-lockdown average pollutant concentration from the pollutant concentrations during lockdown. We compared this with the average values of the pre-lockdown time and converted it into a percentage. We obtained the average percentage of declination due to lockdown for all major activity zones. This can be formulated as follows.

$$\text{AverageDeclination} = \left(\frac{\text{Mean during lockdown} - \text{Mean of the before lockdown}}{\text{Mean of the before lockdown}} \cdot 100\right)\%$$

From Table (7), we can see that in every zone, the parameters are reduced to a certain percentage.

Table 7: Percentage of average decrease of $PM_{2.5}$, $PM_{10}$, and AQI in diferent major-activity zones because of lockdown.

| Components | Transport | Residential | Institutional | Commercial |
|---|---|---|---|---|
| $PM_{2.5}$ | -63.79% | -61.31% | -61.38% | -62.13% |
| $PM_{10}$ | -62.19% | -54.84% | -49.83% | -45.73% |
| AQI | -11.11% | -29.05% | -22.71% | -26.76% |

For example, in the transport zone, the average $PM_{2.5}$ values were reduced by approximately 63.79%, whereas in the transport zone, the average $PM_{10}$ values were reduced by 62.19%, and in the residential zone the average of AQI value declined by almost 29.05% (from Table (7)). Due to the lockdown, the weekly average of $PM_{2.5}$ emissions decreased by 62.15%. The weekly average of $PM_{10}$ is reduced by 53.14%, and the weekly average of AQI decreased by 22.40% throughout Delhi.

### 4.1 Spatial Trend Analysis

#### 4.1.1 Spatial Similarity Analysis

The prime condition for spatial trend analysis is to crosscheck whether the weekly $PM_{2.5}$ emission changes with the geo-spatial location. In this scenario, spatial auto-correlation plays an important role because if the datasets are spatially auto-correlated, then there exists a similar pattern in $PM_{2.5}$, $PM_{10}$ emission, and other pollutant concentrations in the dataset of nearby locations. If the datasets are not spatially auto-correlated for those cases, the datasets are spatially random, which means that the datasets of two nearby locations are contradictory in pattern to each other. Therefore, when studying the spatial trend, we should consider those datasets which are spatially dependent of each other. If the datasets are spatially random, then we cannot interpolate the $PM_{2.5}$ values for the unobserved locations, using the observed information. Therefore, we must first inspect the spatial autocorrelation. Therefore, the test of spatial autocorrelation is executed at a significance level of 0.05, where Moran's $I$ index ($I$) is taken as a measure of spatial autocorrelation (Till 2018). Here,

$$H_0: I = 0$$

$$H_1: I \neq 0$$

In particular, during BL, the data for the $12^{th}$, $23^{rd}$, $25^{th}$, and $26^{th}$ weeks are not spatially auto-correlated, and during DL, the data are not spatially auto-correlated from the $15^{th}$ week to the $19^{th}$ week and from the $24^{th}$ week to the $26^{th}$ week at 0.05 level of significance as the p-value is less than 0.05. Therefore, we can infer that the datasets for the remaining period are spatially autocorrelated. This is why we observe similar types of characteristics in $PM_{2.5}$ emission in neighbouring locations compared to those locations which are geographically far away. For example, the weekly average data of the important air pollutant concentrations, such as $PM_{2.5}$, $PM_{10}$, etc. for the $13^{th}$ week of 2019 are spatially autocorrelated. We studied the patterns of the data and found that Narela is near the monitoring station of Alipur compared to the Karni Singh Shooting Range. As a result, during the $13^{th}$ week of the year 2019 (considered as BL in Table (5)), there was a high correlation between the weekly average of $PM_{2.5}$ in Alipur and Narela (0.98) compared to that of Alipur and Karni Singh Shooting Range (0.93). This test of spatial autocorrelation plays an important role in determining the spatial neighbourhood, which ultimately helps in spatial interpolation in the unobserved location.

**4.1.2 Spatial Interpolation**

Our target was to interpolate the pollutant concentrations at those unobserved locations. In this case, spatial interpolation is a reliable tool. Therefore, for spatial trend analysis resolution, these weekly datasets are taken which are spatially auto-correlated, and the missing data are interpolated using spatial interpolation techniques. In this study, three types of interpolation techniques, IDW, OK, and RFK, were applied to compare the efficiencies of these interpolation techniques. In IDW, the goodness of interpolation is dependent on the value of $p$. We inspected the correctness of the IDW interpolation method with respect to two measures, namely the RMSE and $R^2$ values given in Table (8).

Table 8: RMSE and $R^2$ changing with the variation of $p$ in IDW

| $p$ | RMSE | $R^2$ |
|---|---|---|
| 0.1 | 21.713862 | 0.915102 |
| 0.2 | 21.5611777 | 0.916301 |
| 0.3 | 21.4079816 | 0.9175161 |
| 0.4 | 21.2656589 | 0.9186298 |
| 0.5 | 21.14805 | 0.9589295 |
| 0.6 | 21.0694 | 0.959249 |
| 0.7 | 21.0408 | 0.9593693 |
| 0.8 | 21.06682 | 0.9202081 |
| 0.9 | 21.144 | 0.9196403 |
| 1.0 | 21.264 | 0.9187535 |
| 1.1 | 21.41463 | 0.917632 |
| 1.2 | 21.585 | 0.9163575 |
| 1.3 | 21.76469 | 0.9149997 |
| 1.4 | 21.947072 | 0.9136144 |
| 1.5 | 22.12609 | 0.9122447 |
| 1.6 | 22.29762 | 0.9109228 |
| 1.7 | 22.4589 | 0.9096708 |

| 1.8 | 22.608   | 0.9085022 |
| 1.9 | 22.74565 | 0.9074232 |
| 2.0 | 22.8706  | 0.9064344 |

From Table (8), it can be concluded that the most favourable value of $p$ in this interpolation procedure is 0.7, because the minimum RMSE is 21.0408. In this scenario, the $R^2$ value is 0.9593693 (from Table (8)) which implies that approximately 96% of the $PM_{2.5}$ data is explained by the IDW method, and it is satisfactory to interpolate the unobserved values. Employing the IDW interpolation approach, the gridded points ($.01^0$ by $.01^0$ i.e. 10 km × 10 km along longitude and latitude) were interpolated to show the spatial trend for BL and DL. Despite this, the RMSE of IDW for BL is 16.40 whereas that for DL is 12.72 (from Table (9)).

Table 9: Comparison between IDW, OK, and RFK corresponding to the spatially autocorrelated weeks in 2019 and 2020 on the basis of RMSE.

| Sl.No. | WEEK | YEAR | IDW   | OK    | RFK   |
|--------|------|------|-------|-------|-------|
| 1      | 12   | 2020 | 15.46 | 12.53 | 0.030 |
| 2      | 13   | 2019 | 16.73 | 15.93 | 0.071 |
| 3      | 13   | 2020 | 9.90  | 9.64  | 0.011 |
| 4      | 14   | 2019 | 18.14 | 16.32 | 0.135 |
| 5      | 14   | 2020 | 11.91 | 11.45 | 0.043 |
| 6      | 15   | 2019 | 16.05 | 14.52 | 0.069 |
| 7      | 16   | 2019 | 9.90  | 9.53  | 0.024 |
| 8      | 17   | 2019 | 15.87 | 14.78 | 0.009 |
| 9      | 18   | 2019 | 13.24 | 14.10 | 0.069 |
| 10     | 19   | 2019 | 28.13 | 25.38 | 0.409 |
| 11     | 20   | 2019 | 16.40 | 13.15 | 0.138 |
| 12     | 20   | 2020 | 12.72 | 11.78 | 0.442 |

In OK, the empirical variogram is fitted by several variogram models such as Gaussian, exponential, linear, and spherical, and amidst all of these, the RMSE of the spherical variogram model is minimal, as shown in Table (10) during both periods.

Table 10: Comparison between the performance of different variogram models for $20^{th}$ week during 2019 and 2020.

| S.No | Week | Year | Variogram Model | RMSE         |
|------|------|------|-----------------|--------------|
| 1    | 20   | 2020 | Linear          | 16.50854     |
| 2    | 20   | 2020 | Gaussian        | 56.982       |
| 3    | 20   | 2020 | Spherical       | 0.00170      |
| 4    | 20   | 2020 | Exponential     | 196.3646     |
| 5    | 20   | 2019 | Linear          | 35.33872     |
| 6    | 20   | 2019 | Gaussian        | 149.5183     |
| 7    | 20   | 2019 | Spherical       | 8.940682e-05 |
| 8    | 20   | 2019 | Exponential     | 494.5034     |

From Table (10) we find that the RMSE of the spherical variogram model fitting is very low compared to the other variogram models, such as Gaussian, Linear, and Exponential, to capture the variability information due to spatial reasons. During the $20^{th}$ week of the year 2019, the RMSE was 0.00170 (from Table (10)) which indicates that the spherical variogram is capable of detecting the spatial variability compared to the other variogram models. Before fitting the spherical variogram, the important parameters, nugget, sill, and range were estimated, taking into account the minimum RMSE; for example, the RMSE of the spherical variogram fitting on the dataset of $20^{th}$ week of the year 2019 was 0.0034. Then, using OK, we interpolate the weekly average of $PM_{2.5}$ on the unobserved locations and gridded points. The RMSEs of OK for BL and DL are respectively, 13.15 and 11.78 (from Table (9)).

In the RFK model, at the time of performing the RF regression between the variables to predict the $PM_{2.5}$ values, mtry and ntree were taken differently based on Out-Of-Bag (OOB) errors. For example, the mtry of $20^{th}$ week of 2019 is 4, but for the same week of 2020, the value of mtry is 3, but the ntree is the same, that is, 1000 for both periods. In addition, the value of G (Han 2016) is plotted along the x-axis in Figure (7), and along the y-axis, the names of the variables are plotted. We distinguished that the first three important variables for this RFK model for $20^{th}$ week 2019 were latitude, $PM_{10}$, and longitude, whereas those of the $20^{th}$ week 2020 were $PM_{10}$, CO, and longitude. Therefore, during the $20^{th}$ week of 2020, the variation in $PM_{2.5}$ emission with respect to the variation in latitude, is not very effective.

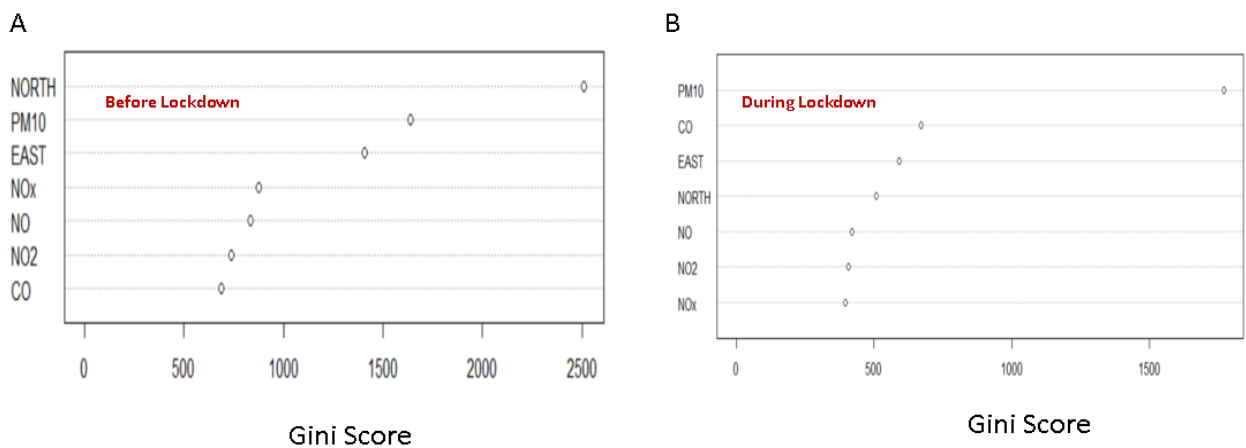

Figure 7: Plotting of Variable importance, that is, Gini-Mean decrease of explanatory variables (used in RFK) in descending order.

From Figure (7 **A**) we can say that latitude is the most important variable in RFK, and in Figure (7 **B**) $PM_{10}$ is the most important variable in the RF regression. Moreover, during the $20^{th}$ week of 2019 and 2020, latitude was positively correlated (0.223), whereas longitude was negatively correlated (-0.33) with $PM_{2.5}$. It implies that if we move along the eastern direction, then pollution decreases with increasing longitude, whereas if we move along the northern direction, the density of $PM_{2.5}$ increases. Therefore, we can infer that the risk of air pollution in northern Delhi is more sensitive than that in the eastern part of Delhi. Thereafter, in RFK, after employing RF regression, the empirical variogram of OOB errors is fitted with a spherical variogram model. The important parameters of the variogram: nugget, sill, range, and the corresponding RMSE are respectively 22.1, 47, 0.28, and 0.013 for the $20^{th}$ week of 2019, whereas for the $20^{th}$ week of 2020, the estimates of those parameters are 3, 24.45, 0.28, respectively and RMSE is 0.0057. In Figure (8 (**A**)) and Figure (8 (**B**)), the spatial trend of BL and DL of $20^{th}$ week is shown as the data for this week is spatially auto-correlated (validated by $I$). In this contour plot, we plotted longitude, latitude, and interpolated $PM_{2.5}$ values along the x-axis, y-axis, and z-axes, respectively.

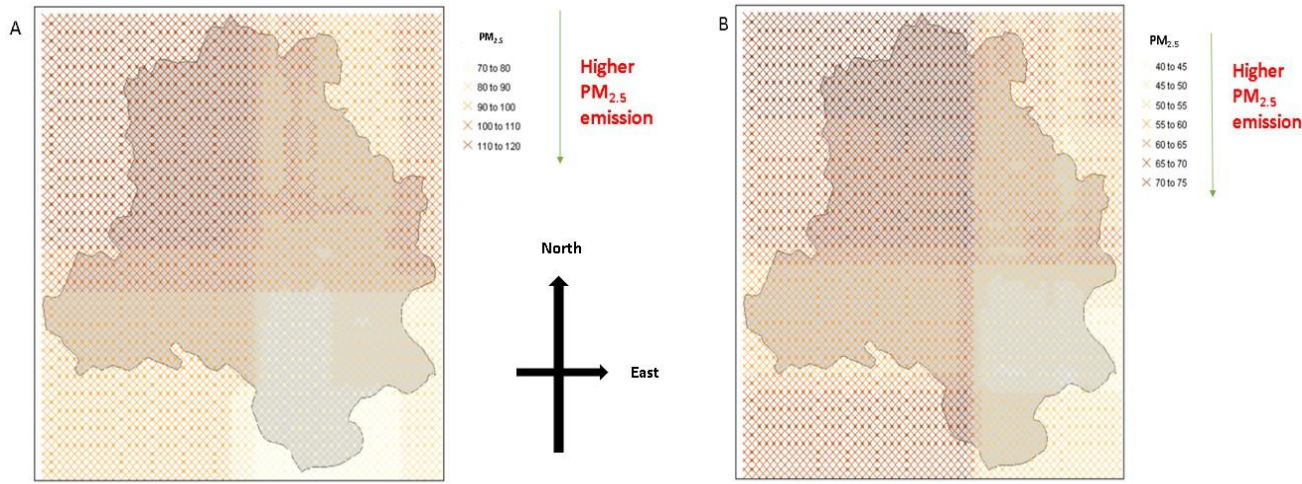

Figure 8: **(A)** Spatial trend for weekly $PM_{2.5}$ emission of $20^{th}$ week for Before lockdown (i.e. the year 2019).

**(B)** Spatial trend for weekly $PM_{2.5}$ emission of $20^{th}$ week for During lockdown (i.e. the year 2020.

**Note:** The green line indicates the weekly higher $PM_{2.5}$ emissions from weekly lower $PM_{2.5}$ emissions.

The interpolated value was calculated for both periods after fitting the spherical variogram of OOB errors. The contour plot in Figure (8) shows the spatial variation in the effect of lockdown across the whole of Delhi. From Figure (8 **(A)**), it is observed that the highest contour line crosses $120 \mu g/m^3$ which means that the maximum weekly $PM_{2.5}$ concentration during the $20^{th}$ week of the year 2019 reaches 120 $\mu g/m^3$. Similarly, the maximum $PM_{2.5}$ density is in $80 - 95 \mu g/m^3$ (from Figure (8 **(A)**)), but from Figure (8 **(B)**), it has been confirmed that the maximum contour line is $65 \mu g/m^3$ and the maximum density is in $50 - 60 \mu g/m^3$ during the $20^{th}$ week of the year 2020. During the pre-lockdown period, in the northwestern part of Delhi, the $PM_{2.5}$ level was higher compared to the other parts of Delhi. In particular, $PM_{2.5}$ emission is the highest in the locations near Bawana, Delhi, whereas in this region during the lockdown period, the $PM_{2.5}$ emission level is much lower than before the lockdown (from Figure (8 **(A)**)). Moreover, Narela, Bawana, Sector 5, Bakoli, Bhaktwarpur, etc. in North Delhi are the main hot spot areas of $PM_{2.5}$ emission. Therefore, the AQI level in the northern part of Delhi was very high at 185. This indicates that people in these regions are at a high risk of acute respiratory problems. The RMSE of the RFK model for the $20^{th}$ week of 2019 was 0.138, whereas that for the same week in 2020 was 0.442. In Table (9), the entire comparison between the three well-known spatial interpolation models and RFK interpolates admirably because of its minimum RMSE. After using RFK interpolation, we estimated the $PM_{2.5}$ values at the unobserved location for both periods.

4.2 Spatio-Temporal behaviour analysis

**4.2.1 Spatio-Temporal Impact of Lockdown**

In this section, we describe the spatio-temporal intervention of the lockdown and the spatio-temporal characteristics of each monitoring station. Using the DID estimator ($\beta_3$) from

Equation (5) we estimated the measure of intervention of lockdown in $PM_{2.5}$ emission in every major activity zone. In Table (11), we studied the regression coefficient of the DID regression, and in the bracket, the standard error of this estimation is discussed.

Table 11: Measurement of the intervention of lockdown using DID estimator.

| Zone | DID Estimator($\beta_3$) | p-value ($H_0: \beta_3 = 0$) |
|---|---|---|
| Transport | -8.13 | 0.00876** |
| Residential | -0.46 | 0.23854. |
| Commercial | -2.78 | 0.039** |
| Institutional | -1.40 | 0.0079* |

**Note:** " *** " for "$p < 0.001$", " ** " for "$p < 0.01$", " * " for "$p < 0.05$", and ". " for "$p < 0.1$".

From Table (11), we observe that in the transport zone, the lockdown has a negative impact (-8.13), meaning that because of the lockdown in this zone, there is a huge restriction regarding transportation, which helps to mitigate the $PM.2.5$ emission. This intervention is significant at 0.1 level, which means that almost 90% of areas under transport zones have a strict declination in $PM_{2.5}$ emission due to lockdown. Likewise, the lockdown also had a significant negative effect on commercial (-2.78) and institutional zones (-1.40), which was significant at 0.1 level, respectively. However, in the residential zone, the lockdown had minimum negative intervention (-0.46), and it was significant at 0.1 level. In the transport zone, this negative influence is the maximum compared with the other zones. Therefore, lockdown in the transport zone is an effective measure for controlling AQI levels.

### 4.2.2 Spatio-Temporal behaviour of Monitoring Stations

In this section, we discuss the spatio-temporal behaviour of each monitoring station to study the trend of $PM_{2.5}$ emissions during the four seasons. As a result, we are able to understand the spatio-temporal features of AQI levels in the neighbourhood of each monitoring station. From Table (12), we detect that in every monitoring station during winter, the MK $\tau$ is positive, which indicates that we should be careful about the air pollution during the winter season except in IHBAS, Dilshad Garden, where the MK $\tau$ is $-0.018$, which is satisfactory because it means that in this region and its neighbourhood the $PM_{2.5}$ emission is low. This helps to maintain healthy AQI. Among all of the monitoring stations and their neighbourhoods, the trend of $PM_{2.5}$ is highly positive; for example, in DTU, the MK $\tau$ is 0.1 and in Dwarka, it is 0.156. Similar statistics are seen in Mundka (0.123), Karni Singh Shooting Range (0.123), and Ayanagar (0.138) which are risky in terms of $PM_{2.5}$ emission (from Table (12)). Similarly, in summer and spring, the MK $\tau$ for almost all monitoring stations were negative. Therefore, the AQI values during the summer and spring were satisfactory and healthy. However, during the monsoon in some monitoring stations and their neighbourhoods, the trend of $PM_{2.5}$ emission is decreasing, and in other monitoring stations, the trend is the opposite. For example, during the monsoon in Burari Crossing (MK $\tau = 0.33$), the air quality is poorer than that of Dwarka (MK $\tau = -0.046$) (from Table (12)). Thus, we analysed the spatio-temporal features of each monitoring station and its neighbourhood.

Table 12: Value of MK $\tau$ for monitoring stations of BL and DL.

| Monitoring Stations | Winter | Spring | Summer | Monsoon |
|---|---|---|---|---|
| Alipur | 0.046 | -0.28 | -0.408 | 0.0312 |
| AnandVihar | 0.069 | -0.28 | -0.39 | -0.035 |
| AshokVihar | 0.068 | -0.35 | -0.49 | 0.068 |
| Ayanagar | 0.138 | -0.30 | -0.17 | 0.114 |
| Bawana | 0.088 | -0.27 | -0.46 | 0.12 |
| Burari Crossing | 0.0018 | -0.26 | -0.43 | 0.33 |
| CRRI mathura road | 0.028 | -0.19 | -0.32 | -0.067 |
| DTU | 0.1 | -0.25 | -0.34 | 0.033 |
| Dwarka | 0.156 | -0.066 | -0.374 | -0.046 |
| IGI Airport | 0.082 | -0.28 | -0.41 | -0.10 |
| IHBAS, Dilshad Garden | -0.018 | -0.27 | -0.301 | -0.29 |
| ITO, Delhi | 0.078 | -0.33 | -0.43 | -0.037 |
| Jahangirpuri, Delhi | 0.09 | -0.28 | -0.44 | -0.00437 |
| Jawaharlal Nehru stadium, Delhi | 0.095 | -0.458 | -0.408 | 0.072 |
| Karni Singh Shooting range | 0.141 | -0.372 | -0.417 | 0.094 |
| Lodhi Road, Delhi | 0.0614 | -0.324 | -0.195 | 0.102 |
| Major Dhyan Chand National Stadium, Delhi | 0.058 | -0.363 | -0.376 | 0.021 |
| Mandir Marg, Delhi | 0.086 | -0.416 | -0.322 | 0.0579 |
| Mundka, Delhi | 0.123 | -0.168 | -0.452 | 0.0465 |
| Najafgarh, Delhi | 0.0584 | -0.247 | -0.469 | 0.0842 |
| Narela, Delhi | 0.0451 | -0.288 | -0.361 | 0.082 |
| Nehru Nagar, Delhi | 0.094 | -0.483 | -0.393 | -0.000547 |
| North Campus, DU, Delhi | 0.053 | -0.347 | -0.463 | -0.0514 |
| NSIT Dwarka, Delhi | 0.11 | -0.121 | -0.371 | -0.054 |
| Okhla Phase | 0.094 | -0.324 | -0.40 | 0.0164 |
| Patparganj, | 0.0894 | -0.378 | -0.247 | -0.011 |

| Location | | | | |
|---|---|---|---|---|
| Delhi | | | | |
| Punjabi Bagh, Delhi | 0.043 | -0.395 | -0.417 | -0.0142 |
| Pusa, Delhi | 0.111 | -0.472 | -0.38 | 0.0536 |
| R K Puram, Delhi | 0.0063 | -0.38 | -0.174 | -0.00984 |
| Rohini, Delhi | 0.0753 | -0.301 | -0.407 | 0.094 |
| Shadipur, Delhi | 0.0504 | -0.348 | -0.371 | -0.0719 |
| Sirifort, Delhi | 0.0889 | -0.299 | -0.42 | 0.169 |
| Sonia Vihar, Delhi | 0.0188 | -0.371 | -0.333 | -0.0437 |
| Sri Aurobindo Marg, Delhi | 0.0897 | -0.397 | -0.36 | 0.0771 |
| Vivek Vihar, Delhi | 0.091 | -0.35 | -0.393 | 0.0361 |
| Wazirpur, Delhi | 0.0648 | -0.445 | -0.424 | 0.161 |

In Table (12), the value of Mann–Kendall's $\tau$ over two time periods for monitoring stations is discussed at a 0.95 level of confidence. We can discern that the temporal trend diminishes almost every time, except for a few monitoring stations, such as in Lodhi road, where the trend is increasing for two time periods, and in Karni Singh, where the trend is increasing in DL.

As a result, the pollution of the entire Delhi region was diluted compared to the previous years. However, RFK performs better than other interpolation models, but some inescapable flaws stipulate further exploration of previous research methodologies, and we admit the restrictions of our research; for example, we have ignored the impact of serious meteorological factors such as wind speed and wind direction on $PM_{2.5}$ emission. However, this type of statistical approach has a significant disadvantage; it does not involve the impact of the changing characteristics of the atmosphere, such as wind speed and direction and depth of the inversion layer.

## 5. Conclusions

This study examined the temporal and spatial patterns of pollutant concentrations throughout the year. We are not only restricted to the important features of pollutant concentration, but we also focus on the spatio-temporal impact of lock-down on air pollution. From the above discussion, we can see that the pollution level of the BL period strongly differs from that of the AL period. There is a huge variation in $PM_{2.5}$ emission in Delhi. During the monsoon season in Delhi, pollution was minimal in the transport zone. In winter, pollution is at its maximum in the transport zone. In addition to seasonal influence, there is a significant secular trend, and cyclic behaviour is observed in 24 hour $PM_{2.5}$ emissions in Delhi. The average pollutant emission in Delhi decreased by approximately 60% because of lockdown, but the lockdown is not the sole factor responsible for this decline. The season also helps decrease the pollutant concentrations in the air. As the daily $PM_{2.5}$ emissions are not completely spatially dependent, there is a serious limitation of the spatio-temporal interpolation techniques for predicting the same. Among the three previously mentioned spatial interpolation techniques, RFK performed better than the other models. Moreover, in the northwestern part of Delhi, the pollution level is serious, and the AQI is very unhealthy, especially in Ayanagar, Dwaraka, Narela, and Bawana Sector 5. In addition to the spatial and temporal cross-sectional discussion about the influence of lockdown, we measured the spatio-temporal intervention of lockdown on pollutant concentration simultaneously, and it indicated that the transport zone was the most affected zone. One of the future directions of this research is to use copulas to find the joint probability distribution function of the covariates, helping us compute the Bayesian risk of wind speed and wind direction of $PM_{2.5}$ concentration. We could not establish a testing procedure for spatial stationarity. Consequently, some portions of the data have an ill-conditioned covariance matrix. Therefore, we can model the data spatially and temporally simultaneously using ML techniques to construct a spatial-temporal model.